\documentclass[twocolumn,floatfix,a4paper,superscriptaddress]{revtex4}
\usepackage{bm,color,amsmath,txfonts}
\usepackage{graphicx}
\usepackage{siunitx}
\usepackage{subfigure}
\usepackage{verbatim}
\usepackage{dcolumn}
\usepackage{bm}
\usepackage{epsf}
\usepackage{xcolor}
\usepackage[colorlinks,
linkcolor=blue,
anchorcolor=blue,
citecolor=blue]{hyperref}
\usepackage{hhline}
\usepackage{braket}
\usepackage{float}
\usepackage{enumerate}
\usepackage{bbm}
\usepackage{lipsum}
\usepackage{mathrsfs}
\usepackage{mathtools}

\begin{document}

\title{Multicritical dissipative phase transitions manipulated by dipole–dipole interactions}
    
\author{Jia-Xin Wang}
\affiliation{College of Physics, Sichuan University, Chengdu 610065, China}
	
\author{Qian Bin\footnote{Contact author: qianbin@scu.edu.cn}}
\affiliation{College of Physics, Sichuan University, Chengdu 610065, China}

\author{Jing-Jing Cao}
\affiliation{College of Physics, Sichuan University, Chengdu 610065, China}

\author{Xin-You L\"{u}\footnote{Contact author: xinyoulu@hust.edu.cn}}
\affiliation{School of Physics and Hubei Key Laboratory of Gravitation and Quantum Physics, Huazhong University of Science and Technology, Wuhan, 430074, P. R. China}

\begin{abstract}
Precise control of criticality in superradiant phase transitions is essential for quantum state engineering and the simulation of nonequilibrium phase transitions. Here, we investigate theoretically multicritical phenomena  in a dissipative two-Rydberg-atom cavity-QED system. The intrinisic dipole--dipole interaction between the two Rydberg atoms restructures the energy-level landscape of the atomic subsystem, thereby significantly modifying the boundary of the continuous second-order superradiant phase transition,  and shifting both the phase boundary and the multicritical point toward weaker atom–cavity strengths.  For sufficiently strong dipole--dipole interactions, the continuous second-order phase transition and the multicritical point both disappear, leaving only a discontinuous first-order phase transition that enables the emergence of a superradiant phase even at arbitrarily weak atom--cavity coupling. This work is of fundamental interest for studying dissipative quantum phase transitions, with potential implications for quantum precision measurement and quantum sensing.

\end{abstract}
\maketitle

\section{Introduction}
	
Superradiance, first proposed by Dicke in 1954, refers to the collective enhancement of spontaneous emission from an ensemble of two-level atoms coupled to a cavity mode~\cite{1}. This cooperative phenomenon has found broad applications in superradiant lasers~\cite{2,3}, superradiance lattices~\cite{4}, quantum refrigerator~\cite{5,6}, black‑hole superradiance~\cite{7,8}, and superradiant scattering~\cite{9,10}. More fundamentally, the Dicke model itself also hosts a superradiant phase transition~\cite{11,12,13}. In the thermodynamic limit (\(N \to \infty\)), as the atom-field coupling strength exceeds a critical threshold, the system undergoes a phase transition from a normal phase with vanishing photon population to a superradiant phase with a macroscopic coherent cavity field. The transition is driven by spontaneous breaking of a residual \(\mathbb{Z}_2\)  symmetry, which arises from the reduction of the original  \(U(1)\) symmetry due to the presence of counter-rotating terms~\cite{14,15}.  Beyond this fundamental superradiant phase transition,  multicritical superradiant phase transitions, where distinct phase boundaries (e.g., first-order and second-order) intersect at a multicritical point, have recently attracted considerable attention~\cite{16,17}. Near such points, the system exhibits extreme sensitivity to parameter variations~\cite{18,19}, diverging susceptibility, and nontrivial crossover scaling behaviors~\cite{20,21}. These features make multicritical phase transitions potential valuable for quantum sensing and metrology, where an enhanced response to external perturbations is highly desirable~\cite{22,23}. At the multicritical point, the merging of first- and second-order transition lines provides an additional control dimension,  allowing the sensor to operate in either a continuous-response mode or a threshold-detection mode within the same device~\cite{24}.

In recent years, Rydberg atoms have attracted considerable attention in quantum information processing and quantum simulation owing to their strong dipole moments, long lifetimes, and high degree of controllability~\cite{25,26,27,28}. These unique properties have enabled the exploration of a wide range of quantum phenomena, including coherent excitation transfer~\cite{29,30}, time crystals~\cite{31,32,33}, many-body localization~\cite{34,35}, quantum scars~\cite{36,37}, false-vacuum decay and bubble nucleation~\cite{38}, and topological phase~\cite{39}.
Furthermore, it is these same large dipole moments and long lifetimes that make Rydberg atoms exceptionally well suited for strong coupling to resonant cavities. Rydberg atoms coupled to resonant cavities thus provide a versatile cavity-QED platform for studying light–matter interactions~\cite{40,41,42,43,44,45,46,47,48}. Such systems have been employed to investigate a variety of collective phenomena, including multicritical behavior~\cite{16,17,49}, local and nonlocal dynamics~\cite{50}, superradiant clock phases~\cite{51}, and confined meson excitations~\cite{52}.  However, the vast majority of these studies have focused exclusively on atom–cavity interactions, treating the atoms as non-interacting entities. The intrinsic dipole–dipole interactions between Rydberg atoms, by contrast, offer a complementary and equally important degree of freedom.  Owing to their large dipole moments, such interactions are readily realizable and can become a dominant energy scale that profoundly reshapes the collective behavior~\cite{53}. Exploring how they modify collective phenomena may uncover novel phase structures with implications for quantum sensing and metrology.

Here, we investigate multicritical dissipative phase transitions in a cavity-QED system consisting of two Rydberg atoms with intrinsic dipole–dipole interactions.  Using a mean-field approximation supplemented by quantum fluctuation analysis~\cite{54,55,56,57}, we show that the dipole–dipole interaction fundamentally modifies the critical behavior of the superradiant phase transition. In particular, it can continuously modify the phase boundaries of both the first- and second-order superradiant transitions, shifting the second-order phase boundary and the associated multicritical point toward weaker atom--cavity coupling strengths. For sufficiently strong dipole–dipole interactions, the second-order phase transition and the associated multicritical point are eventually disappear, leaving only a discontinuous first-order phase transition that enables the emergence of a superradiant phase even at arbitrarily weak atom-cavity coupling.  This behavior originates from the dipole–dipole interaction effectively restructuring the energy-level landscape of the atomic subsystem and thereby altering the critical condition for the second-order phase transition. To further characterize the different phases, we compute the Wigner function directly from the quantum master equation. Its phase-space distribution provides clear signatures of the normal phase, the superradiant phase, and the bistable coexistence region. These results establish a connection between intrinsic dipole–dipole interactions and the engineering of multicriticality, with promising implications for criticality-enhanced quantum sensing~\cite{58,59,60,61,62}.

\section{MODEL}

\begin{figure}[t]
	\centering
	\includegraphics[width=8.5cm]{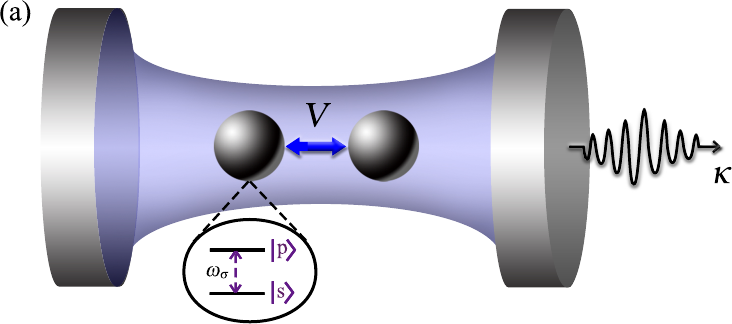}%
	\vspace{0.3cm}   %
	\includegraphics[width=8.7cm]{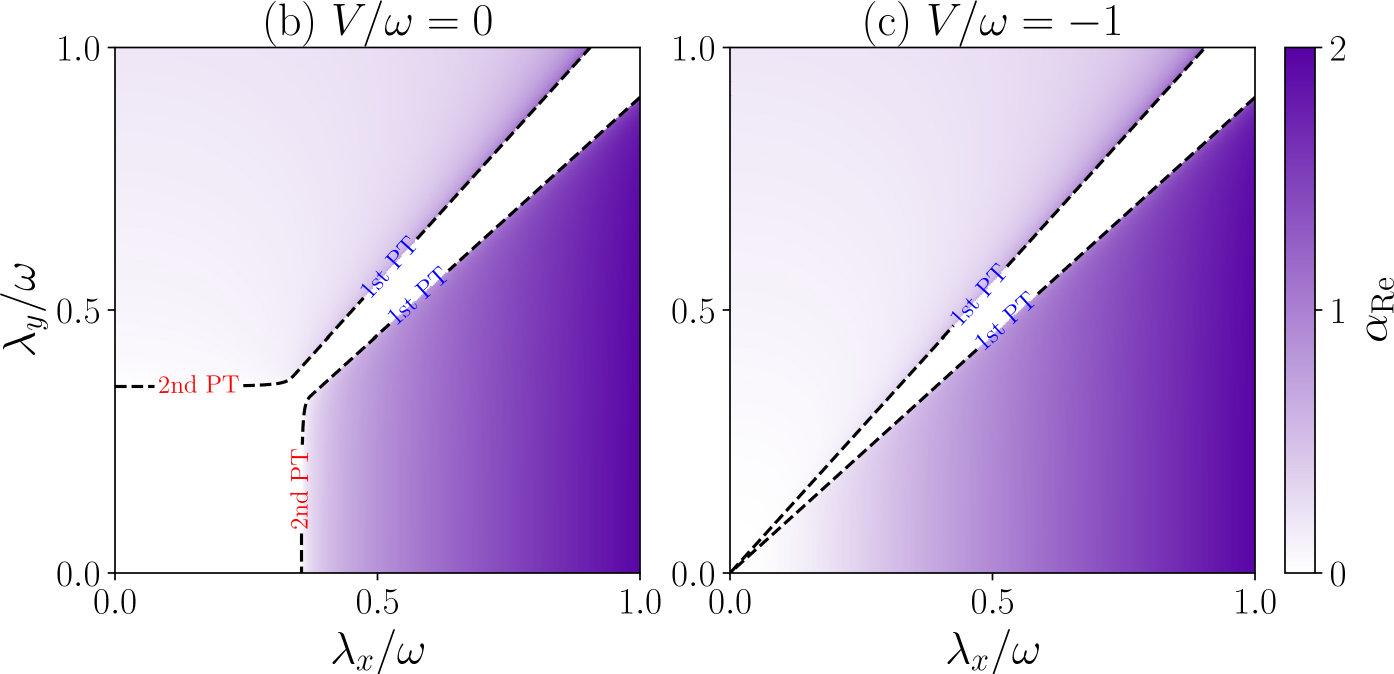}
	\caption{(a) Schematic of the model. Two identical Rydberg atoms with levels \(|\mathrm{s}\rangle\) and \(|\mathrm{p}\rangle\) are placed inside a cavity. Here \(V\) denotes the strength of the dipole--dipole interaction between the two atoms, and \(\kappa\) is the cavity decay rate.  (b, c) The real part of cavity field amplitude $\alpha$ as a function of the coupling strengths $\lambda_x$ and $\lambda_y$ for (b) \(V/\omega = 0\) and (c) \(V/\omega = -1\), respectively. The labels ``1st PT" and ``2nd PT" denote first-order and  second-order phase transitions. System parameters: \(\omega_c = \omega_{\sigma_j} = \omega\) and \(\kappa/\omega = 0.1\).}
	\label{1}
\end{figure}

We consider a cavity QED system consisting of a single cavity mode coupled to two interacting Rydberg atoms, as schematically illustrated in Fig.~\ref{1}(a). The atoms are modeled as effective two-level systems, each with a ground state \(|\mathrm{s}\rangle\) and a first excited state \(|\mathrm{p}\rangle\). In addition to the atom--cavity coupling, the two atoms interact via a dipole--dipole interaction of strength \(V = C_3 / R^3\), where \(C_3\) is the dipole--dipole coupling parameter and \(R\) is the interatomic distance. Since \(V \propto1/R^3\), the interaction strength can be readily tuned by adjusting the atomic separation using optical tweezers~\cite{63}. The total Hamiltonian of the system thus reads (\(\hbar = 1\))
\begin{align}
	\label{e1}
	\begin{split}
		H = &\ \omega_c a^\dagger a + \frac{1}{2} \sum_{j=1}^{2} \omega_{\sigma_j} \sigma_j^z + {V} \bigl( \sigma_1^+ \sigma_2^- + \sigma_1^- \sigma_2^+ \bigr) \\
		&+ \sum_{j=1}^{2} \Bigl[ \lambda_{xj} \bigl(a + a^\dagger\bigr) \sigma_j^x + i \lambda_{yj} \bigl(a - a^\dagger\bigr) \sigma_j^y \Bigr].
	\end{split}
\end{align}
The first two terms represent the free energies of the cavity mode and the two atoms, respectively. The cavity mode with frequency \(\omega_c\) is described by the bosonic creation and annihilation operators \(a^\dagger\) and \(a\), while the two atoms with frequencies \(\omega_{\sigma_1}\) and \(\omega_{\sigma_2}\) are represented by the Pauli spin operators \(\sigma_j^\alpha\) (\(\alpha = x,y,z\), \(j = 1,2\)). The third term describes the dipole--dipole interaction between the two Rydberg atoms, where \(\sigma_j^\pm=(\sigma_j^x\pm i\sigma_j^y)/2 \) denote the atomic raising and lowering operators. This flip--flop interaction originates from dipole--dipole coupling between highly excited Rydberg states and mediates excitation exchange between the two atoms.
The fourth term describes the coupling between the two quadratures of the cavity mode and the two atoms, with coupling strengths \(\lambda_{xj}\) and \(\lambda_{yj}\), respectively. Microscopically, this interaction arises from the electric dipole coupling \(-\hat{\bm{d}} \cdot \hat{\bm{E}}\), where the atomic dipole operator \(\hat{\bm{d}}\)  and the quantized cavity field can be decomposes into the \(\sigma^x\), \(\sigma^y\) and  \((a + a^\dagger)\), \(i(a - a^\dagger)\) quadratures, respectively.

The symmetry of the system is governed by the coupling strengths $\lambda_{xj}$ and $\lambda_{yj}$. When \(\lambda_{xj} = \lambda_{yj}\), the Hamiltonian possesses a continuous \(U(1)\) symmetry under the transformation \(a \to a e^{i\theta}\) and \(\sigma_j^- \to \sigma_j^- e^{i\theta}\), corresponding to the two-atom Tavis--Cummings model~\cite{64}. When $\lambda_{xj} \neq \lambda_{yj}$, this continuous symmetry is broken and only two independent parity symmetries remain, giving rise to a  $\mathbb{Z}_2 \times \mathbb{Z}_2$ symmetry. The first $\mathbb{Z}_2$ transformation is $\sigma_j^x \to -\sigma_j^x$ together with $a + a^\dagger \to -(a + a^\dagger)$ ($\alpha_{\text{Re}} \to -\alpha_{\text{Re}}$), while the second is $\sigma_j^y \to -\sigma_j^y$ together with $i(a - a^\dagger) \to -i(a - a^\dagger)$ ($\alpha_{\text{Im}} \to -\alpha_{\text{Im}}$). In the limiting case where one of the coupling vanishes, the model reduces to the two-atom Dicke model and retains only a single $\mathbb{Z}_2$ symmetry~\cite{65}. 

Since superradiant phase transitions are associated with spontaneous symmetry breaking, the symmetry structure of the Hamiltonian plays a crucial role in determining the phase diagram of the system. In the presence of cavity dissipation, the system exhibits nonequilibrium superradiant phase transitions. For weak atom--cavity couplings, the system remains in the normal phase with a vanishing cavity-field amplitude $\alpha = 0$, as shown in Fig.~\ref{1}(b). Above the critical boundary, the  \(\mathbb{Z}_2\) symmetry is spontaneously broken, giving rise to a superradiant phase characterized by $\alpha \neq 0$~\cite{16,17}. Interestingly, when the two orthogonal couplings strengths $\lambda_{xj} = \lambda_{yj}$ are nearly equal, a narrow region of the normal phase persists along the diagonal direction of the phase diagram, even though each individual coupling is strong. This leads to multicritical behavior~\cite{16,17}. Next, we will discuss in detail the multicritical dissipative phase transition.

\section{MULTICRITICAL DISSIPATIVE PHASE TRANSITIONS}
In this section, we discuss the multicritical behavior of the superradiant phase transition in the two-Rydberg-atom cavity-QED system. First, we derive the mean-field equations of motion and map out the steady-state phase diagram. We then examine the critical behavior of the phase transition and verify the stability of the system through a quantum fluctuation analysis. Finally, based on the full quantum master equation, we compute the Wigner function of the cavity field to characterize the phase transitions in phase space.

\subsection{Mean-Field solutions}

In general, the system inevitably interacts with its surrounding environment, and  its dynamics can be described within the Heisenberg--Langevin formalism. For any system operator \(O\), the equation of motion is governed by
\begin{equation}\label{e2}
	\dot{O} = -i[O, H_{\text{sys}}] - \gamma_O O + \sqrt{2\gamma_O}\,O_{\text{in}}(t),
\end{equation}
where \(H_{\text{sys}}\) is the system Hamiltonian, \(\gamma_O\) denotes the dissipation rate associated with \(O\), and \(O_{\text{in}}(t)\) is the corresponding input noise operator. The first term on the right‑hand side describes the coherent evolution driven by \(H\), the second term accounts for dissipation induced by the environment, and the last term represents the quantum noise entering the system. For the cavity mode \(a\), the dissipation rate is given by the cavity decay rate \(\kappa\). Throughout this work, the spontaneous emission of the Rydberg atoms is neglected, since it is much smaller than the cavity decay rate  (\(\gamma \ll \kappa\)).

From Eq.~\eqref{e2}, we obtain the Heisenberg--Langevin equations of motion for the cavity and atomic operators 
\begin{subequations}\label{e3}
	\begin{align}
		& \begin{aligned}
			\dot{a} &= -(i\omega_c + \kappa) a - i\lambda_{x1}\sigma_1^x - i\lambda_{x2}\sigma_2^x - \lambda_{y1}\sigma_1^y \\
			&\quad - \lambda_{y2}\sigma_2^y + \sqrt{2\kappa}\,a_{\text{in}}(t),
		\end{aligned} \label{e3a}\\
		& \dot{\sigma}_1^x = -\omega_{\sigma1}\sigma_1^y + 2i\lambda_{y1}(a - a^\dagger)\sigma_1^z + V\sigma_1^z\sigma_2^y, \label{e3b}\\
		& \dot{\sigma}_1^y = \omega_{\sigma1}\sigma_1^x - 2\lambda_{x1}(a + a^\dagger)\sigma_1^z - V\sigma_1^z\sigma_2^x, \label{e3c}\\
		& \dot{\sigma}_2^x = -\omega_{\sigma2}\sigma_2^y + 2i\lambda_{y2}(a - a^\dagger)\sigma_2^z + V\sigma_2^z\sigma_1^y, \label{e3d}\\
		& \dot{\sigma}_2^y = \omega_{\sigma2}\sigma_2^x - 2\lambda_{x2}(a + a^\dagger)\sigma_2^z - V\sigma_2^z\sigma_1^x. \label{e3e}
	\end{align}
\end{subequations}
These equations can be separated into their mean values and quantum fluctuations through the decomposition \(O = \langle O \rangle + \delta O\), where \(\langle O \rangle\) denotes the expectation value of the operator and \(\delta O\) represents the corresponding fluctuation operator. Within the mean-field approximation, quantum fluctuations are neglected and the expectation value of a product of operators are factorized into the product of their individual expectation values, e.g., \(\langle a \sigma_1^z \rangle \approx \langle a \rangle \langle \sigma_1^z \rangle\). Setting the time derivatives to zero, we then obtain a closed set of steady-state mean-field equations,
\begin{subequations}\label{e4}
\begin{align}
	& \omega_c \alpha_{\text{Im}} - \kappa \alpha_{\text{Re}} - \lambda_{y1} Y_1 - \lambda_{y2} Y_2 = 0, \label{e4a}\\
	& \omega_c \alpha_{\text{Re}} + \kappa \alpha_{\text{Im}} + \lambda_{x1} X_1 + \lambda_{x2} X_2 = 0, \label{e4b}\\
	& \omega_{\sigma_1} Y_1 + 4\lambda_{y1} \alpha_{\text{Im}} Z_1 - V Z_1 Y_2 = 0, \label{e4c}\\
	& \omega_{\sigma_1} X_1 - 4\lambda_{x1} \alpha_{\text{Re}} Z_1 - V Z_1 X_2 = 0, \label{e4d}\\
	& \omega_{\sigma_2} Y_2 + 4\lambda_{y2} \alpha_{\text{Im}} Z_2 - V Z_2 Y_1 = 0, \label{e4e}\\
	& \omega_{\sigma_2} X_2 - 4\lambda_{x2} \alpha_{\text{Re}} Z_2 - V Z_2 X_1 = 0, \label{e4f}
\end{align}
\end{subequations}
where \(\langle \sigma_j^x \rangle = X_j\), \(\langle \sigma_j^y \rangle = Y_j\), \(\langle \sigma_j^z \rangle = Z_j\), and the expectation value \(\langle a \rangle\) = $\alpha$ with $\alpha = \alpha_{Re} + i \alpha_{Im}$. 

By solving Eqs.~\eqref{e4c}--\eqref{e4f}, we obtains the relations for \(X_j\) and \(Y_j\) of the two atoms
\begin{subequations}\label{e5}
\begin{align}
	X_1 &= \frac{4\alpha_{\text{Re}} Z_1 (\lambda_{x1}\omega_{\sigma_2} + \lambda_{x2} V Z_2)}{\omega_{\sigma_1} \omega_{\sigma_2} - V^2 Z_1 Z_2}, \label{e5a}\\
	X_2 &= \frac{4\alpha_{\text{Re}} Z_2 (\lambda_{x2}\omega_{\sigma_1} + \lambda_{x1} V Z_1)}{\omega_{\sigma_1} \omega_{\sigma_2} - V^2 Z_1 Z_2}, \label{e5b}\\
	Y_1 &= -\frac{4\alpha_{\text{Im}} Z_1 (\lambda_{y1}\omega_{\sigma_2} + \lambda_{y2} V Z_2)}{\omega_{\sigma_1} \omega_{\sigma_2} - V^2 Z_1 Z_2}, \label{e5c}\\
	Y_2 &= -\frac{4\alpha_{\text{Im}} Z_2 (\lambda_{y2}\omega_{\sigma_1} + \lambda_{y1} V Z_1)}{\omega_{\sigma_1} \omega_{\sigma_2} - V^2 Z_1 Z_2}. \label{e5d}
\end{align}
\end{subequations}
Since we consider two identical atoms, the parameters for both atoms are taken to be the same, $\omega_{\sigma 1} = \omega_{\sigma 2} = \omega$ and $\lambda_{x1}=\lambda_{x2} = \lambda_x, \lambda_{y1}=\lambda_{y2} = \lambda_y$. Substituting Eqs.~\eqref{e5a}--\eqref{e5d} into the cavity equations, Eqs.~\eqref{e4a} and \eqref{e4b}, yields a quartic equation for the atomic polarization \(Z\)
\begin{align}\label{e6}
	&P_4 Z^4 + P_3 Z^3 + P_2 Z^2 + P_1 Z + P_0 = 0.\\
	\shortintertext{where the coefficients are given by}
	&P_0 = (\omega_c^2 + \kappa^2)\omega^4,\nonumber\\
	&P_1 = 8\omega_c\omega^3(\lambda_x^2 + \lambda_y^2),\nonumber\\
	&P_2 = -2(\omega_c^2 + \kappa^2)\omega^2 V^2 + 8\omega_c\omega^2 V(\lambda_x^2 + \lambda_y^2) + 64\omega^2\lambda_x^2\lambda_y^2,\nonumber\\
	&P_3 = -8\omega_c\omega V^2(\lambda_x^2 + \lambda_y^2) + 128\omega V\lambda_x^2\lambda_y^2,\nonumber\\
	&P_4 = (\omega_c^2 + \kappa^2)V^4 - 8\omega_c V^3(\lambda_x^2 + \lambda_y^2) + 64V^2\lambda_x^2\lambda_y^2,\nonumber
\end{align}
whose physical solutions completely determine the steady-state properties of the system. Note that the atom expectation value \(\langle \sigma^z \rangle\) lies in the range \([-1, 1]\), with \(Z = -1\) in the normal phase and $Z \in (-1, 0]$ in the superradiant phase.  When multiple physical solutions exist, the dynamically stable steady-state solution is selected, as confirmed by the stability analysis presented below.

Combining the above equations with the spin-conservation relation \(X^2 + Y^2 + Z^2 = 1\), we obtain the cavity-field amplitudes 
\begin{subequations}\label{e7}
\begin{align}
	\alpha_{\text{Re}} &= \pm \sqrt{\frac{(1 - Z^2)(\omega^2 - V^2 Z^2)^2}{16Z^2(\omega + VZ)^2(\lambda_x^2 + R^2\lambda_y^2)}}, \\
	\alpha_{\text{Im}} &= R \alpha_{\text{Re}},
\end{align}
\end{subequations}
where $R = \frac{\kappa(\omega^2 - V^2 Z^2)}{\omega_c(\omega^2 - V^2 Z^2) + 8\lambda_y^2 Z(\omega + VZ)}$. From Eq.~\eqref{e7} one sees that in the normal phase the symmetry is conserved and the cavity field amplitude $\alpha = 0$. In the superradiant phase, in contrast,  two symmetry-related steady-state solutions with opposite cavity fields emerge, reflecting the spontaneous breaking of the \(\mathbb{Z}_2\) symmetry and giving rise to $\alpha \neq 0$.

Figures~\ref{1}(b) and \ref{1}(c) show the  real part \(\alpha_{\text{Re}}\) of $\alpha$ as functions of the coupling strengths \(\lambda_{x}\) and \(\lambda_{y}\) for different values of the dipole--dipole interaction strength \(V\). The blank region corresponds to the normal phase with \(\alpha_{\text{Re}} = 0\), whereas the colored region with \(\alpha_{\text{Re}} \neq 0\) signals the emergence of the superradiant phase. The black dashed lines mark the phase boundaries between the normal phase and the superradiant phase.  For \(V/\omega = 0\), indicated in Fig.~\ref{1}(b), the phase diagram contains both discontinuous first-order (1st PT)  and continuous second-order superradiant  (2nd PT)  phase transitions. When the dipole--dipole interaction is attractive, as shown in Fig.~\ref{1}(c) where \(V/\omega = -1\), the 2nd PT boundary gradually shrinks and eventually disappears, leaving only the 1st PT. Consequently, the superradiant phase can emerge even at arbitrarily weak atom--cavity coupling strengths. The critical behavior of the second-order phase transition can be understood from the mean-field critical coupling condition 
\begin{equation}\label{e8}
	\lambda_y = \sqrt{\frac{(\omega_{\sigma} + V) \left[ ( \kappa^2 + \omega_c^2)(\omega_{\sigma} + V) - 8\omega_c\lambda_x^2 \right]}{8[\omega_c(\omega_{\sigma} + V) - 8\lambda_x^2]}},
\end{equation}
which is obtained from the mean-field equations evaluated at \(Z = -1\). For \(\lambda_x = 0\) this expression reduces to \(\lambda_{c} = \sqrt{(\kappa^2 + \omega_c^2)(\omega_\sigma + V)/(8\omega_c)}\), which explicitly shows that the dipole--dipole interaction \(V\) renormalizes the effective atomic frequency from \(\omega_\sigma\) to \(\omega_\sigma + V\), thereby shifting the critical coupling of the second-order superradiant phase transition. As the attractive interaction becomes stronger, the factor \(\omega_\sigma + V\) decrease. When \(V\rightarrow-\omega_\sigma\), the critical coupling approaches zero, whereas for \(V<-\omega_\sigma\) it becomes imaginary, indicating that the second-order phase boundary ceases to exist. Consequently, only the first-order transition survives in this regime. By contrast, repulsive dipole--dipole interactions (\(V > 0\)) shift the second-order phase boundary toward larger atom--cavity coupling strengths, i.e, the upper right corner in Fig.~\ref{1}(b).

\begin{figure}[t]
	\centering
	\includegraphics[width=8.5cm]{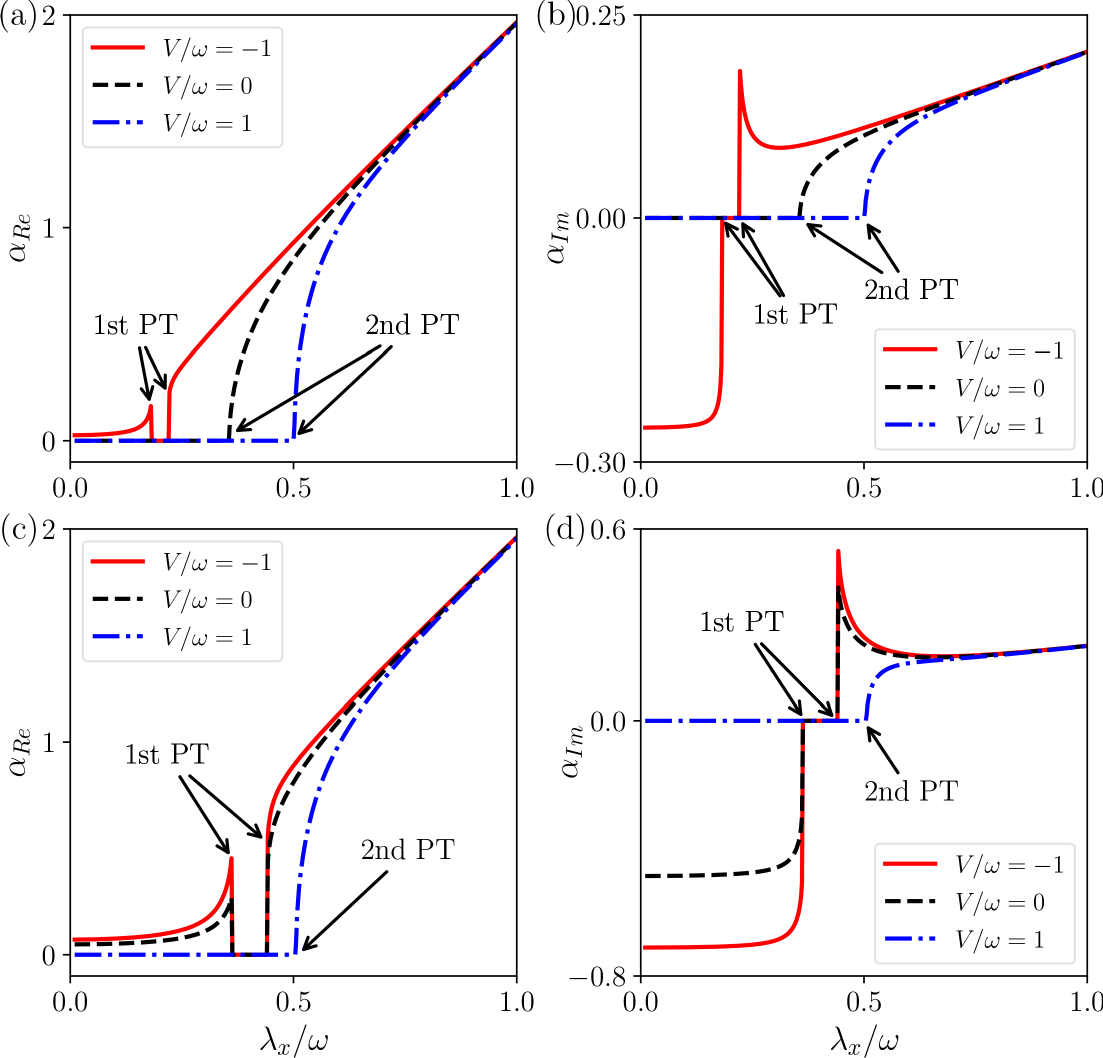}
	\caption{ (a, c) Real part \(\alpha_{\text{Re}}\) and (b,d) imaginary part \(\alpha_{\text{Im}}\)  of the cavity field amplitude $\alpha$ versus  \(\lambda_x\) for different  \(V/\omega\). Red solid, black dashed, and blue dotted curves correspond to  \(V/\omega = -1\), \(V/\omega = 0\) , and \(V/\omega = 1\), respectively. Black arrows mark the positions of the 1st PT and 2nd PT. System parameters:  \(\lambda_y/\omega = 0.2\) for (a,b), \(\lambda_y/\omega = 0.4\) for (c,d), and the other parameters are the same as those in Figs.~\ref{1}(b,c).}
	\label{2}
\end{figure}

To further elucidate the characteristics of the superradiant phase transition, we present in Fig.~\ref{2} the real parts \(\alpha_{\text{Re}}\) and imaginary parts \(\alpha_{\text{Im}}\) of the cavity-field amplitude \(\alpha\) as functions of \(\lambda_x/\omega\) for different dipole--dipole interaction strengths \(V/\omega\), where \(\omega_c = \omega_{\sigma_j} = \omega\). The order parameter exhibits two qualitatively distinct critical behaviors, with a discontinuous jump across a first-order transition and a continuous onset across a second-order transition. The upper panels, Figs.~\ref{2}(a) and \ref{2}(b), correspond to a fixed coupling strength \(\lambda_y/\omega = 0.2\), while the lower panels, Figs.~\ref{2}(c) and \ref{2}(d), are obtained with \(\lambda_y/\omega = 0.4\). The real part \(\alpha_{\text{Re}}\) is displayed in Figs.~\ref{2}(a) and \ref{2}(c), and the imaginary part \(\alpha_{\text{Im}}\) in Figs.~\ref{2}(b) and \ref{2}(d). In each panel, three curves are plotted, corresponding to \(V/\omega = -1\) (red solid line), \(V/\omega = 0\) (black dashed line), and \(V/\omega = 1\) (blue dotted line). Black arrows explicitly mark the locations of the first-order phase transitions characterized by a discontinuous jump and the second-order phase transitions characterized by a continuous change. From these curves two main conclusions can be drawn. First, for fixed \(\lambda_y/\omega\) a change of the dipole--dipole interaction from \(V/\omega = -1\) to \(V/\omega = 1\) continuously modifies the phase boundaries and changes the nature of the superradiant phase transition from first order to second order, as seen by comparing the curves within each panel of Fig.~\ref{2}(a)--\ref{2}(d). Second, comparing the upper and lower sets of panels reveals that a similar change in the order of the phase transition can be induced by varying \(\lambda_y/\omega\) at fixed \(V/\omega\). For instance, when dipole--dipole interaction strength \(V/\omega = 0\), the continuous second-order phase transition for \(\lambda_y/\omega = 0.2\) (upper panels) transforms into a discontinuous first-order phase transition for \(\lambda_y/\omega = 0.4\) (lower panels). These results demonstrate that both the intrinsic dipole--dipole interaction strength and the relative magnitude of the coupling parameters \(\lambda_x/\omega\) and \(\lambda_y/\omega\) offer versatile control for the order of the superradiant phase transition.

\subsection{Quantum fluctuation analysis}

To further clarify how the dipole--dipole interaction influences the critical behavior of the dissipative phase transition, we analyze the quantum fluctuations around the mean-field steady state by linearizing the quantum Langevin equations.

We first introduce the quadrature operators of the cavity-field fluctuations and the corresponding input noise operators as
\begin{subequations}\label{e9}
\begin{align}
	\delta \alpha_{\text{Re}} &= \frac{\delta a + \delta a^\dagger}{2}, \qquad 
	\delta \alpha_{\text{Im}} = \frac{\delta a - \delta a^\dagger}{2i}, \\
	\delta \alpha_{\text{Re}}^{\text{in}} &= \frac{\delta a_{\text{in}} + \delta a_{\text{in}}^{\dagger}}{2}, \qquad 
	\delta \alpha_{\text{Im}}^{\text{in}} = \frac{\delta a_{\text{in}} - \delta a_{\text{in}}^{\dagger}}{2i}.
\end{align}
\end{subequations}
The cavity input noise operators satisfy the correlation functions~\cite{66}
\begin{subequations}\label{e10}
\begin{align}
	\langle a^{\mathrm{in} \dagger}(t') a^{\mathrm{in}}(t) \rangle &= 0, \\
	\langle a^{\mathrm{in}}(t) a^{\mathrm{in} \dagger}(t') \rangle &= \delta(t - t'),
\end{align}
\end{subequations}
where the Markov approximation has been assumed.

To close the set of equations, we eliminate the fluctuation \(\delta Z\) of the \(z\)-component of the spin by linearizing the spin conservation constraint \(X_{j}^2 + Y_{j}^2 + Z_{j}^2 = 1\) around the mean-field values, which gives
\begin{align}\label{e11}
	\delta z_{j} = -\frac{X_{j0} \delta x_{j} + Y_{j0} \delta y_{j}}{Z_{j0}}.
\end{align}
 Substituting the quadrature definitions Eqs.~\eqref{e9}  and the above relation Eqs.~\eqref{e11} into  equations Eqs.~\eqref{e3}, and introducing the input noise operators associated with the cavity dissipation, we obtain a closed set of linearized equations of motion
\begin{subequations}\label{e12}
\begin{align}
	\delta \dot{\alpha}_{\text{Re}} &= -\kappa \delta \alpha_{\text{Re}} + \omega_c \delta \alpha_{\text{Im}} - \lambda_{y1} \delta y_1 - \lambda_{y2} \delta y_2 + \sqrt{2\kappa} \delta \alpha_{\text{Re}}^{\text{in}}, \\
	\delta \dot{\alpha}_{\text{Im}} &= -\omega_c \delta \alpha_{\text{Re}} - \kappa \delta \alpha_{\text{Im}} - \lambda_{x1} \delta x_1 - \lambda_{x2} \delta x_2 + \sqrt{2\kappa} \delta \alpha_{\text{Im}}^{\text{in}}, \\
	\delta \dot{x}_1 &= -4\lambda_{y1} Z_{10} \delta \alpha_{\text{Im}} + \frac{X_{10}}{Z_{10}} \bigl(4\lambda_{y1}\alpha_{\text{Im}} - VY_{20}\bigr) \delta x_1 + V Z_{10} \delta y_2 \notag\\
	&\quad + \left( -\omega_{\sigma_1} + \frac{Y_{10}}{Z_{10}} \bigl(4\lambda_{y1}\alpha_{\text{Im}} - VY_{20}\bigr) \right) \delta y_1, \\
	\delta \dot{y}_1 &= -4\lambda_{x1} Z_{10} \delta \alpha_{\text{Re}} + \left( \omega_{\sigma_1} + \frac{X_{10}}{Z_{10}} \bigl(4\lambda_{x1}\alpha_{\text{Re}} + VX_{20}\bigr) \right) \delta x_1 \notag\\
	&\quad + \frac{Y_{10}}{Z_{10}} \bigl(4\lambda_{x1}\alpha_{\text{Re}} + VX_{20}\bigr) \delta y_1 - V Z_{10} \delta x_2, \\
	\delta \dot{x}_2 &= -4\lambda_{y2} Z_{20} \delta \alpha_{\text{Im}} + V Z_{20} \delta y_1 + \frac{X_{20}}{Z_{20}} \bigl(4\lambda_{y2}\alpha_{\text{Im}} - VY_{10}\bigr) \delta x_2 \notag\\
	&\quad + \left( -\omega_{\sigma_2} + \frac{Y_{20}}{Z_{20}} \bigl(4\lambda_{y2}\alpha_{\text{Im}} - VY_{10}\bigr) \right) \delta y_2, \\
	\delta \dot{y}_2 &= -4\lambda_{x2} Z_{20} \delta \alpha_{\text{Re}} - V Z_{20} \delta x_1 + \frac{Y_{20}}{Z_{20}} \bigl(4\lambda_{x2}\alpha_{\text{Re}} + VX_{10}\bigr) \delta y_2 \notag\\
	&\quad + \left( \omega_{\sigma_2} + \frac{X_{20}}{Z_{20}} \bigl(4\lambda_{x2}\alpha_{\text{Re}} + VX_{10}\bigr) \right) \delta x_2.
\end{align}
\end{subequations}
The set of linearized quantum Langevin equations can be cast into a compact matrix form
\begin{align}\label{e13}
	\dot{u}(t) = M u(t) + \nu(t),
\end{align}
with the fluctuation vector defined as \(u(t) = [\delta \alpha_{\text{Re}}, \delta \alpha_{\text{Im}}, \delta x_{1}, \delta y_{1}, \delta x_{2}, \delta y_{2}]^{T}\) and the corresponding noise operator vector \(\nu(t) = [\sqrt{2\kappa} \delta \alpha_{\text{Re}}^{\text{in}}, \sqrt{2\kappa} \delta \alpha_{\text{Im}}^{\text{in}}, 0, 0, 0, 0]^{T}\). For identical atoms ($X_{10} = X_{20}=X, Y_{10} = Y_{20}=Y, Z_{10} = Z_{20}=Z$) and identical atom-cavity couplings $\lambda_{x1}=\lambda_{x1}=\lambda_{x}, \lambda_{y1}=\lambda_{y1}=\lambda_{y}$, the coefficient matrix \(M\) that governs the linearized dynamics takes the form
\begin{align}\label{e14}
	\setlength{\arraycolsep}{1pt}
	M = \begin{bmatrix}		
		-\kappa & \omega_c & 0 & -\lambda_{y} & 0 & -\lambda_{y} \\[4pt]
		-\omega_c & -\kappa & -\lambda_{x} & 0 & -\lambda_{x} & 0 \\[4pt]
		0 & -4\lambda_{y}Z & \frac{X}{Z}A & -\omega + \frac{Y}{Z}A  & 0 & V Z \\[4pt]
		-4\lambda_{x}Z & 0 & \omega+\frac{X}{Z}B & \frac{Y}{Z}B & -V Z & 0 \\[4pt]
		0 & -4\lambda_{y}Z & 0 & V Z & \frac{X}{Z}A & -\omega+\frac{Y}{Z} \\[4pt]
		-4\lambda_{x}Z & 0 & -V Z & 0 & \omega+\frac{X}{Z}B & \frac{Y}{Z}B
	\end{bmatrix},
\end{align}
where the auxiliary quantities \(A\) and \(B\) are defined as $A = 4\lambda_{y}\alpha_{\text{Im}} - V Y, B = 4\lambda_{x}\alpha_{\text{Re}} + V X$. The stability of each mean-field solution is determined from the eigenvalues of the drift matrix.  According to the Routh--Hurwitz criterion, a steady state is dynamically stable only when all eigenvalues of  the coefficient matrix \(M\) have negative real parts~\cite{67}. We apply the criterion to every mean-field solution and to the quantum fluctuation analysis, retaining only those that are dynamically stable.

Due to the linearized dynamics and the Gaussian property of the quantum noises, the state of the system is fully characterized by the \(6 \times 6\) covariance matrix \(V\), with the matrix elements defined as
\begin{align}\label{15}
	V_{ij}(\infty) = \frac{1}{2} \bigl[ \langle u_{i}(\infty) u_{j}(\infty) \rangle + \langle u_{j}(\infty) u_{i}(\infty) \rangle \bigr].
\end{align}
In the steady-state regime, the covariance matrix can be determined directly by solving the standard Lyapunov equation~\cite{68}
\begin{align}\label{e16}
	M V + V M^{T} = -D,
\end{align}
where $D=\mathrm{diag}[\kappa/2, \kappa/2, 0, 0, 0, 0]$ is the diffusion matrix, which is defined through $D_{ij}\delta(t-t^{\prime})=\langle \nu_i (t)\nu_j (t^{\prime})+\nu_j (t^{\prime})\nu_i (t)\rangle/2$. Within this framework, the variance of the cavity mode quadrature, which quantifies the photon number fluctuations, is expressed in terms of the covariance matrix elements as $\langle \delta \alpha^\dagger \delta \alpha \rangle = V_{11} + V_{22} - 1 / 2$. Notably, the procedure for calculating photon number fluctuations here is equivalent to employing the Holstein-Primakoff transformation~\cite{69,70}.

\begin{figure}[t]
	\centering
	\includegraphics[width=8.7cm]{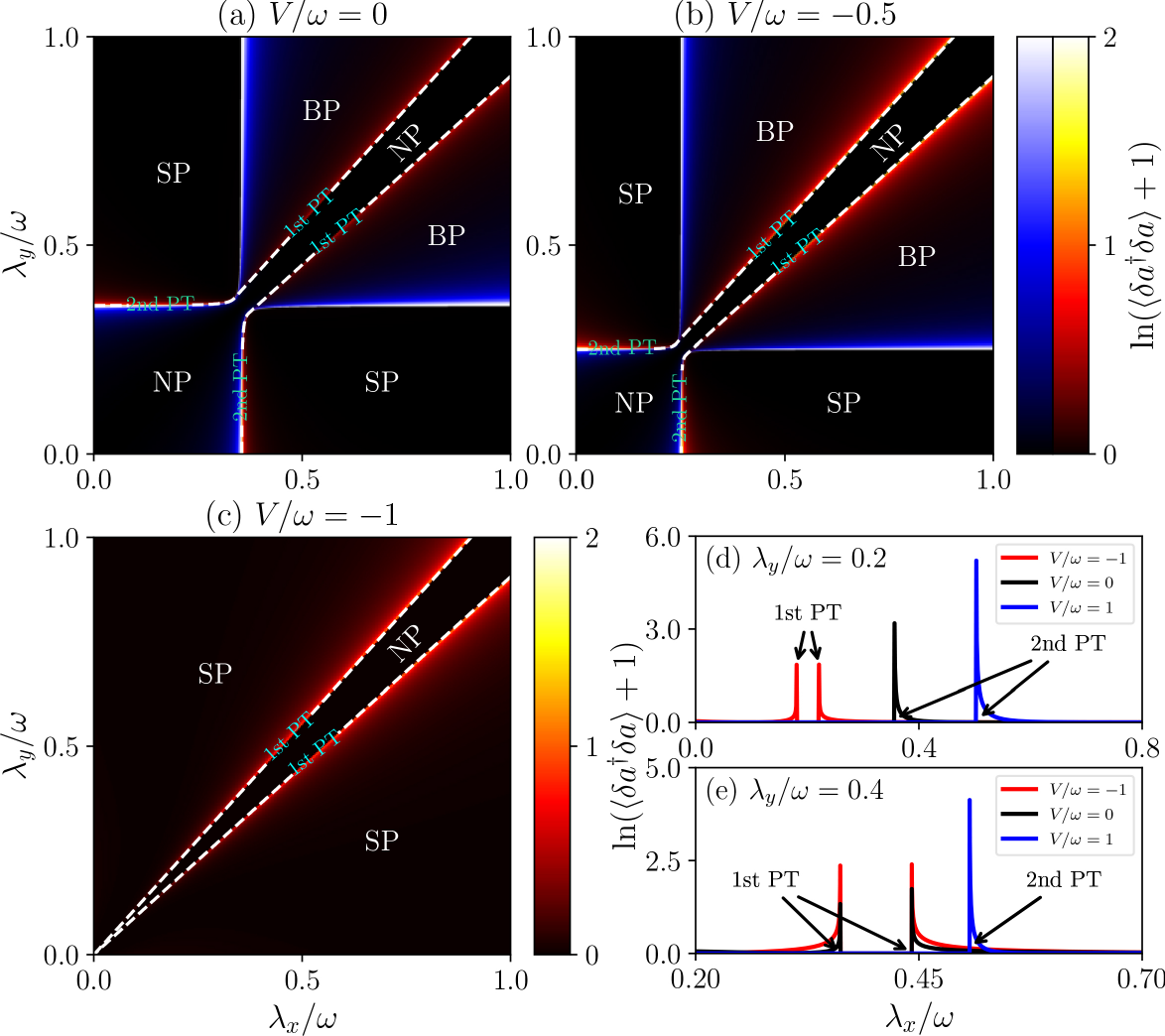}
	\caption{Quantum  fluctuations of the cavity field  \(\ln(\langle \delta \alpha^\dagger \delta \alpha \rangle + 1)\) obtained from the mean‑field solutions as functions  \(\lambda_{x}/\omega\) and \(\lambda_{y}/\omega\) for (a) \(V/\omega = 0\), (b) \(V/\omega = -0.5\), and (c) \(V/\omega = -1\). Blue and red regions correspond to the normal and superradiant phases, respectively, and their overlap indicating the bistable coexistence region. The labels ``NP'', ``SP'', and ``BP'' mark pure normal, pure superradiant, and bistable phases. Panels (d) and (e) show the fluctuations as a function of \(\lambda_{x}/\omega\) for fixed \(\lambda_{y}/\omega = 0.2\) and \(\lambda_{y}/\omega = 0.4\), respectively. The other parameters are the same as those in Figs.~\ref{1}(b,c).}
	\label{3}
\end{figure}

Figure~\ref{3} show the quantum fluctuations of the cavity field calculated from the mean-field steady states. The quantity \(\ln(\langle \delta \alpha^\dagger \delta \alpha \rangle + 1)\) is adopted to characterize the fluctuations as a function of the various system parameters. Figures~\ref{3}(a), \ref{3}(b), and \ref{3}(c) show the fluctuations as functions of the coupling strengths \(\lambda_{x}/\omega\) and \(\lambda_{y}/\omega\) for different dipole--dipole interaction strengths, \(V/\omega = 0\), \(V/\omega = -0.5\), and \(V/\omega = -1\). The blue and red regions correspond to the fluctuations of the normal phase and the superradiant phase, respectively. The domains labeled ``NP'', ``SP'', and ``BP'' in the figure denote the pure normal phases, pure superradiant phases, and the bistable coexistence region where both phases can be simultaneously stable.  Specifically, the system stays in the normal phase with both atoms in their ground states when \(\lambda_x < \lambda_c\) and \(\lambda_y < \lambda_c\) or \(\lambda_x \approx \lambda_y\). If one coupling exceeds \(\lambda_c\) and the other remains below it the system enters the superradiant phase. For \(\lambda_x > \lambda_c\) and \(\lambda_y > \lambda_c\) with a sufficiently large difference the system becomes bistable, with the normal and superradiant solutions coexisting stably. From these three panels we observe that when the dipole--dipole interaction is attractive (\(V/\omega < 0\)), the critical line of the continuous second‑order phase transition shifts toward the lower left corner of the phase diagram and eventually disappears, leaving only the discontinuous first‑order transition boundary. This is in full agreement with the evolution of the real part \(\alpha_{\text{Re}}\) of the cavity field obtained from the mean-field calculation discussed above. It should be emphasized here that when \(V/\omega = -1\), the fluctuation of the normal phase vanishes from the phase diagram,  and only the superradiant phase survives. The fundamental reason is that the eigenvalues of the coefficient matrix \(M\) of the normal phase are zero at this point, making the normal phase fluctuation unstable and therefore vanish, and the photon-number fluctuations in that regime originate entirely from the superradiant branch. 

To show this feature more clearly,  Figs.~\ref{3}(d) and \ref{3}(e) show two‑dimensional cuts of the fluctuations \(\ln(\langle \delta \alpha^\dagger \delta \alpha \rangle + 1)\) as a function of a single coupling strength \(\lambda_{x}/\omega\). Panels (c) and (d) correspond to \(\lambda_{y}/\omega = 0.2\) and \(\lambda_{y}/\omega = 0.4\), respectively, with different curves in each panel represent different dipole-dipole interaction strengths \(V/\omega\). The fluctuation behavior closely follows the evolution of the order parameter shown in Fig.~\ref{2}. In particular, tuning the dipole--dipole interaction continuously modifies the phase boundaries and changes the character of the superradiant phase transition from discontinuous first order to continuous second order. These results further demonstrate that intrinsic dipole--dipole interaction has a profound influence on the critical properties of the quantum phase transition and can provide an effective means for their manipulation.

\subsection{The Wigner function}

\begin{figure}[t]
	\centering
	\includegraphics[width=7.5cm]{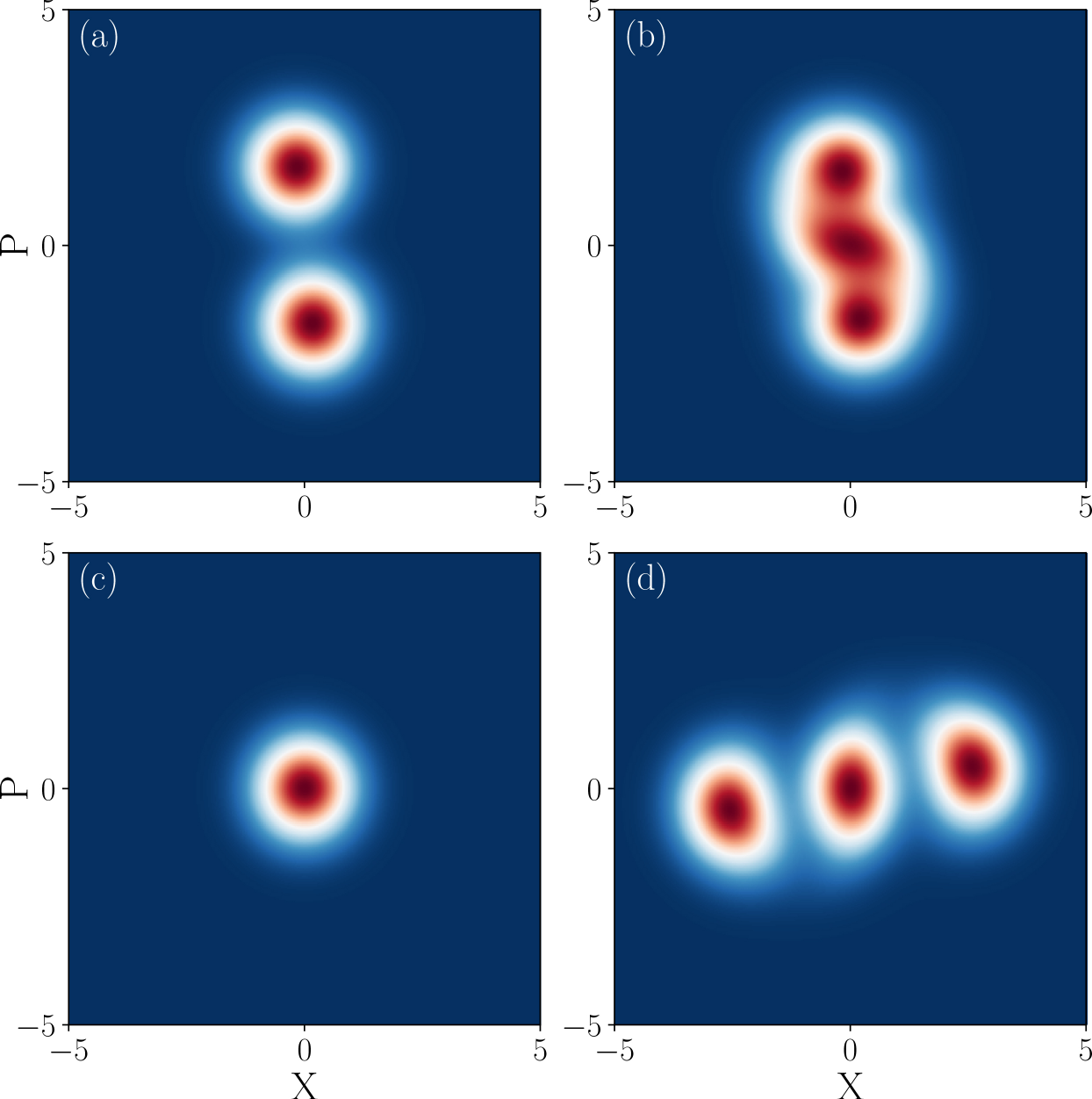}
	\caption{Cavity-field Wigner function for \(V/\omega = -0.5\) at fixed \(\lambda_{y}/\omega = 0.62\) and (a) \(\lambda_{x}/\omega = 0.1\), (b) \(\lambda_{x}/\omega = 0.388\), (c) \(\lambda_{x}/\omega = 0.65\), (d) \(\lambda_{x}/\omega = 0.95\). The quadrature components of the cavity field fluctuations are denoted by \(X = \sqrt{2}\delta \alpha_{\text{Re}}\) and \(P = \sqrt{2}\delta \alpha_{\text{Im}}\).The other parameters are the same as those in Figs.~\ref{1}(b,c).}
	\label{4}
\end{figure}

To further characterize the steady-state phases beyond the mean-field and linearized fluctuation analyses, we calculate the cavity-field Wigner function directly from the full quantum master equation. Unlike the mean-field approximation, this approach fully incorporates quantum fluctuations and therefore provides an intuitive phase-space representation of the steady state. Since the spontaneous emission rate of the Rydberg atoms is negligible compared with the cavity decay rate, we set $\gamma = 0$ throughout, leading to
\begin{equation}\label{e17}
	\dot{\rho} = -i [H, \rho] + \kappa \mathcal{D}[a] \rho,
\end{equation}
where the dissipator takes the standard Lindblad form \(\mathcal{D}[a]\rho = 2 a \rho a^\dagger - a^\dagger a \rho - \rho a^\dagger a\).
The steady-state density matrix obtained from Eq.~\eqref{e17} is used to construct the cavity-field Wigner function~\cite{71}
\begin{align}\label{e18} 
	W(\alpha) = \frac{1}{\pi^2} \int d^2 \beta \, C_W(\beta) \, e^{\beta^* \alpha - \beta \alpha^*},
\end{align}
where \(\alpha\) denotes the complex amplitude in phase space, \(C_W(\beta) = \operatorname{Tr}(\rho D(\beta))\) is the characteristic function, and \(D(\beta) = \exp(\beta a^\dagger - \beta^* a)\) stands for the displacement operator. The Wigner function provides a direct visualization of the cavity-field state in phase space and therefore offers an intuitive characterization of the different dissipative phases.

Figure~\ref{4} shows the numerically calculated Wigner functions for several representative parameter sets at a fixed dipole--dipole interaction strength \(V/\omega = -0.5\). It should be noted that the present system contains only two Rydberg atoms, so finite-size effects are significant. Consequently, the sharp critical behavior predicted by the mean-field theory is partially smoothed in the full quantum solution, and the Wigner function does not exhibit perfectly sharp phase boundaries. Nevertheless, it clearly captures the characteristic phase-space signatures of the different steady-state phases and agrees well with the phase diagrams obtained from the mean-field and quantum fluctuation analyses. To illustrate the evolution of the phase-space distribution across the phase diagram, four representative points are selected along the line \(\lambda_y/\omega=0.62\) in Fig.~\ref{3}(b). Figure~\ref{4}(a) corresponds to the regime where only \(\lambda_y\) exceeds the critical coupling. In this case, the Wigner function exhibits two symmetric peaks displaced along the \(Y\)-quadrature direction, indicating the spontaneous breaking of the \(\mathbb{Z}_2\) symmetry and the formation of the superradiant phase. As \(\lambda_x\) is increased such that both coupling strengths exceed the critical value while \(\lambda_x<\lambda_y\), the system enters the bistable coexistence region. As shown in Fig.~\ref{4}(b), the Wigner function develops a characteristic three-peak structure consisting of a central peak located at the origin and two symmetric side peaks. The central peak represents the normal phase, whereas the side peaks correspond to the superradiant phase, demonstrating that both steady states coexist and are simultaneously stable. 

When the two coupling strengths become nearly equal \(\lambda_x\approx\lambda_y\), the system re-enters the narrow normal-phase region. Accordingly, the two side peaks disappear and only a single peak centered at the origin remains, as shown in Fig.~\ref{4}(c). Increasing the coupling further to the regime \(\lambda_x>\lambda_y\) drives the system back into the bistable region. Consequently, the three-peak structure reappears in Fig.~\ref{4}(d). In contrast to Fig.~\ref{4}(b), however, the two side peaks are now displaced along the \(X\)-quadrature direction rather than the \(Y\)-quadrature direction. The rotation of the side peaks in phase space directly reflects the competition between the two orthogonal atom--cavity couplings. When \(\lambda_y\) dominates, the superradiant order primarily develops in the \(Y\)-quadrature of the cavity field, whereas for \(\lambda_x>\lambda_y\) it is transferred to the \(X\)-quadrature. More importantly, the emergence and disappearance of the three-peak structure provide direct evidence for the appearance of the bistable coexistence region predicted by the mean-field theory. The Wigner-function analysis therefore offers a full-quantum confirmation of the multicritical dissipative phase diagram and reveals the phase-space signatures associated with the normal phase, the superradiant phase, and their coexistence.

\section{CONCLUSION}

In summary, we have investigated multicritical dissipative phase transitions in a two-Rydberg-atom cavity-QED system with intrinsic dipole–dipole interactions. By combining a mean-field approximation with a detailed quantum fluctuation analysis, we have demonstrated that the dipole--dipole interaction significantly modifies the superradiant phase diagram by continuously tuning the phase boundaries. In particular, it shifts the second-order phase boundary toward weaker atom--cavity coupling strengths and eventually suppresses it completely, leaving only a first-order phase transition. As a consequence, the superradiant phase can emerge even at arbitrarily weak atom--cavity coupling, a regime inaccessible in the conventional Dicke model. These critical features are further confirmed by the cavity-field Wigner function obtained from the full quantum master equation, whose distinct single-, double-, and triple-peak structures provide clear signatures of the normal phase, the superradiant phase, and the bistable coexistence region. Our results demonstrate that intrinsic dipole--dipole interactions provide a efficient means for engineering multicritical dissipative phase transitions and suggest that similar interaction-driven critical phenomena may also arise in other cavity-QED and many-body quantum systems, with potential applications in criticality-enhanced quantum metrology and sensing.

\section*{ACKNOWLEDGMENTS}

This work is supported by  the National Science Fund for Distinguished Young Scholars of China (Grant No.\,12425502),  the National Natural Science Foundation of China (Grants  No.\,12574397 and No.\,12547108), the Quantum Science and Technology-National Science and Technology Major Project (Grant No.\,2024ZD0301000), the National Key Research and Development Program of China (Grant No.\,2021YFA1400700), the Sichuan Science and Technology Program (Grant No.\,2025ZNSFSC0057).

\bibliography{basename of .bib file}

\end{document}